\documentclass[12pt]{article}
\usepackage{graphicx}
\usepackage{siunitx}

\newcommand\pubdate{\today}

\def\Title#1{\begin{center} {\Large #1 } \end{center}}
\def\Author#1{\begin{center}{ \sc #1} \end{center}}
\def\Address#1{\begin{center}{ \it #1} \end{center}}

\newcommand\pubblock{\rightline{\begin{tabular}{l}  \\ 
         \pubdate  \end{tabular}}}
\newenvironment{Abstract}{\begin{quotation}  }{\end{quotation}}
\newenvironment{Presented}{\begin{quotation} \begin{center} 
             PRESENTED AT\end{center}\bigskip 
      \begin{center}\begin{large}}{\end{large}\end{center} \end{quotation}}

\begin{document}
\begin{titlepage}
 \pubblock
\vfill
\Title{The Scattering and Neutrino Detector at the LHC}
\vfill
\Author{A. Iuliano (On behalf of the SND@LHC Collaboration)}
\Address{Università di Napoli Federico II and INFN}
\vfill
\begin{Abstract}
SND@LHC is a compact and stand-alone experiment to perform measurements with neutrinos produced at the LHC in a hitherto unexplored pseudo-rapidity region of $7.2 < \eta < 8.4$, complementary to all the other experiments at the LHC. The experiment is located \SI{480}{\meter} downstream of ATLAS IP1 in the unused TI18 tunnel. The detector is composed of a hybrid system based on an 800 kg target mass of tungsten plates, interleaved with emulsion and electronic trackers, followed downstream by a calorimeter and a muon system. The configuration allows efficiently distinguishing between all three neutrino flavours, opening a unique opportunity to probe physics of heavy flavour production at the LHC in the region that is not accessible to ATLAS, CMS and LHCb. This region is of particular interest also for future circular colliders and for predictions of very high-energy atmospheric neutrinos. The detector concept is also well suited to searching for Feebly Interacting Particles via signatures of scattering in the detector target. The first phase aims at operating the detector throughout LHC Run 3 to collect a total of $250 \mbox{fb}^{-1}$. This presentation reports the current status of the analysis, after the data collected in the first year of data taking. A new era of collider neutrino physics is just starting. 
\end{Abstract}
\vfill
\begin{Presented}
DIS2023: XXX International Workshop on Deep-Inelastic Scattering and
Related Subjects, \\
Michigan State University, USA, 27-31 March 2023 \\
\end{Presented}
\vfill
\end{titlepage}
%


\section{Introduction and Motivation for SND@LHC}

The LHC (Large Hadron Collider) is currently the highest energy collider in the world. It provides high energy neutrinos, allowing to scope the currently unexplored domain in the $O(10^2\,\mbox{GeV})$ - $O(1\,\mbox{TeV})$ range. Measuring neutrinos of all the flavours can fill the gap between accelerator measurements and data from cosmic rays. The SND@LHC (Scattering and Neutrino Detector at the Large Hadron Collider) aims to provide these measurements, identifying all three neutrino flavours.

The physics program of the experiment covers the following subjects:

\begin{itemize}
	\item The measurement of the $\sigma_{pp \rightarrow \nu X}$ cross section: $\nu_e$ and $\nu_\tau$ come mainly from charmed hadron decays, while $\nu_\mu$ spectra contain a soft component from $\pi$ and $K$ decay;
	\item Charm production in proton collisions: The off-axis location of the detector, in the pseudorapidity range $7.2 < \eta < 8.4$, focuses on the measurement of the charmed sources;
	\item Gluon PDF Measurement: the data collected by the experiment will provide invaluable input for future colliders and the study of astrophysical sources;
	\item Lepton flavour universality: obtained from cross section ratio measurements in neutrino interactions. The ratio between electron and tau neutrinos depends only on charm hadronization and decay branching fractions;
	\item Measurement of $\sigma_{NC}/\sigma_{CC}$: the identification of the charged leptons separates Charged Current(CC) neutrino interactions from Neutral Current (NC) neutrino interactions. The ratio of the observed events can be written as a function of the Weinberg angle, providing a control measurement for the physical accuracy of the experiment.
\end{itemize}

\begin{figure}[htbp]
	\centering
	\includegraphics[width=0.5\textwidth]{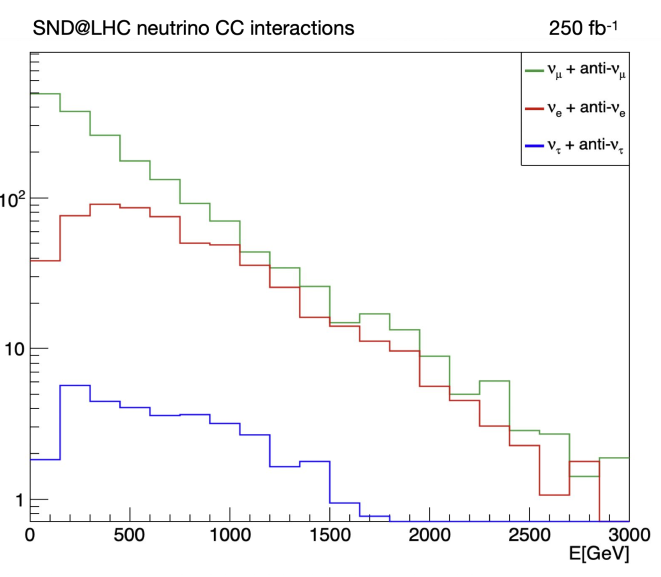}
	\caption{Neutrino spectras, after the integration of $250\,\mbox{fb}^{-1}$.}\label{fig:neutrino_yields}
\end{figure}

The experiment will collect, during Run 3 of the LHC, a total integrated luminosity of \SI{250}{\femto \barn ^{-1}}. In total, $\sim$ 1690 Charged Current and $\sim$ 555 Neutral Current neutrino interactions are expected. The spectra of the neutrinos, obtained with Monte Carlo simulations, are shown in Figure~\ref{fig:neutrino_yields}.

\section{Detector layout}
\begin{figure}[htbp]
	\centering
	\includegraphics[width=0.7\textwidth]{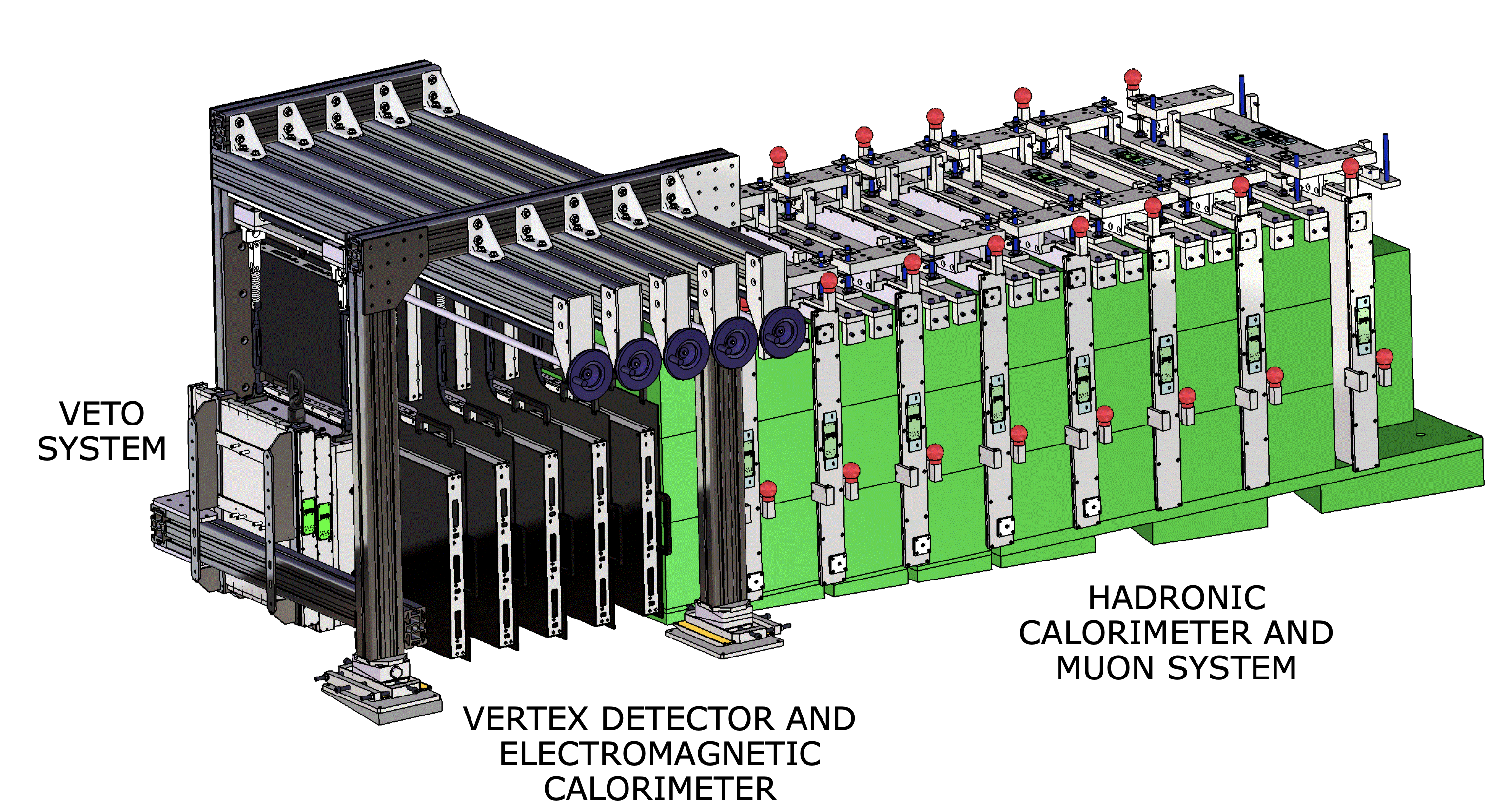}
	\caption{Layout of the SND@LHC detector, showing the three main components~\cite{detector_paper}.}\label{fig:detector_layout}
\end{figure}

The SND@LHC experiment is located at \SI{480}{\meter} from the collision point of the ATLAS experiment (IP1), in the TI18 tunnel of the LHC. The detector is composed of three main sections: a veto system, to tag penetrating muons, a vertex detector employing ECC (Emulsion Cloud Chambers) and SciFi(Scintillating Fibers), also acting as electromagnetic calorimeter, and a muon identification system and hadronic calorimeter downstream. The layout is shown in Figure~\ref{fig:detector_layout}.

\begin{figure}[htbp]
	\centering
	\includegraphics[width=0.9\textwidth]{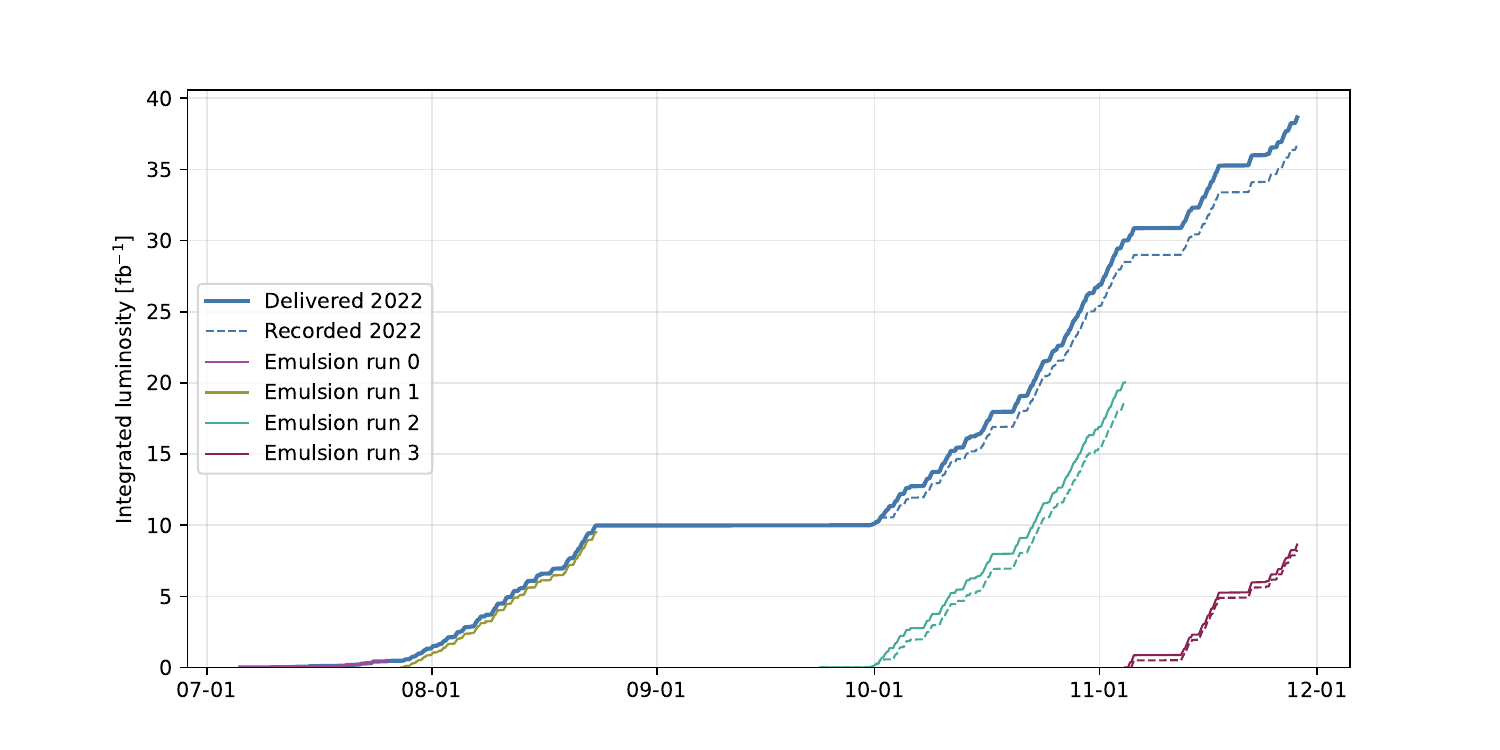}
	\caption{Total delivered and recorded integrated luminosity in the 2022 year. The amount of the single emulsion targets is also shown.}~\label{fig:luminosity_plot}
\end{figure}

After the approval of the experiment in March 2021, the detector has been installed in one year, in order to be ready for the start of Run 3 in April 2022. During the 2022 data taking, a total integrated luminosity of \SI{38.7}{\femto \barn ^{-1}} has been delivered, with the exposure of 4 emulsion targets. The luminosity increase with time is reported in Figure~\ref{fig:luminosity_plot}.

\section{Status of the analysis}

The first physics measurement performed has been the muon flux, aimed to validate Monte Carlo simulations and to evaluate the background for the next neutrino searches. The electronic detectors~\cite{muon_flux} and the nuclear emulsions resulted in consistent measurements of the muon flux, amounting to $(1.5 \pm 0.1)\times 10^4 \mbox{\,fb}/\mbox{\,cm}^2$ in the brick closest to the Line Of Sight in the transverse plane. After updating the geometry of the LHC magnets, consistent yields have been estimated with Monte Carlo simulations.

A search for high energy $\nu_\mu$ CCDIS interactions has been performed using all the data taken by the electronic detectors in 2022. This analysis has been performed adopting a selection with a strong rejection power, designed to yield a clean set of events. First, a fiducial region in the target has been considered, including only events located in the 3rd or 4th wall in order to enhance rejection power for muon-induced background, also ensuring that neutrino-induced showers are sampled by at least two SciFi planes. Then, the event is identified as signal after a cut-based procedure, requiring large hadronic activity in the calorimeter system, a clean outgoing muon track reconstructed in the muon system, and hit time distribution consistent with IP1 direction as source. 

\begin{figure}[htbp]
	\centering
	\includegraphics[width=0.7\textwidth]{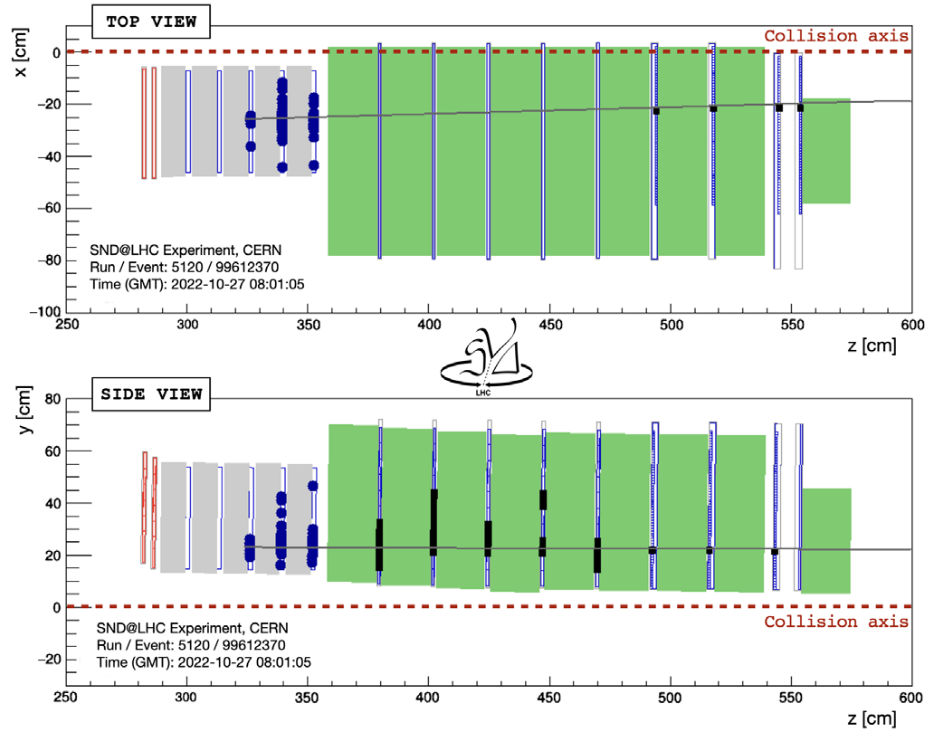}
	\caption{Display of a $\nu_\mu$ CC candidate event. Hits in the SciFi are shown as blue markers, while hits in the hadron calorimeter and muon system are shown as black bars. The line represents the reconstructed muon track\cite{neutrino_paper}.}~\label{fig:neutrino_candidate}
\end{figure}

The main source of background are muons reaching the detector location. The muons can either enter the fiducial volume without being vetoed, or interact in the surrounding material and produce neutral particles that can then mimic neutrino interactions in the target. The background has been estimated by measuring the muon flux, the veto inefficiency, amounting to $4.5 \times 10^{-4}$ during the 2022 run, and estimating the rate of neutral hadron events in muon DIS simulations.
A total of 8 candidate events, consistent with $\nu_\mu$ CC interactions, has been observed. The total amount of background, including both muon and neutral hadrons sources, has been estimated to $(8.6 \pm 3.8) \times 10^{-2}$ events, which implies an excess of $\nu_\mu$ CC signal over the background-only hypothesis of 6.8 standard deviations~\cite{neutrino_paper}. One of the selected candidates is shown in Figure~\ref{fig:neutrino_candidate}. 

\section{Advanced SND@LHC detector}

\begin{figure}[htbp]
	\centering
	\includegraphics[width=0.7\textwidth]{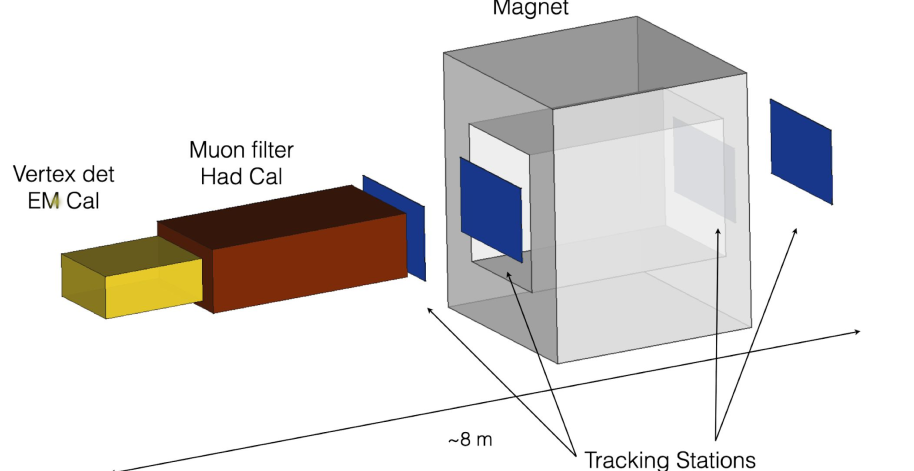}
	\caption{Schematic drawing of the upgraded SND@LHC detector for HL-LHC.}\label{fig:AdvSND_layout}
\end{figure}

Along with data taking and analysis, the SND@LHC collaboration is planning the upgrade for the upcoming High Luminosity Large Hadron Collider (HL-LHC). The current "Far" detector will be completely renovated (Figure~\ref{fig:AdvSND_layout}), replacing nuclear emulsions with an electronic vertex detector, more suited to handle the increased beam luminosity. A magnetic spectrometer will be added, allowing to measure the charge of produced muons, thus separating muon neutrino CC interactions from anti-neutrinos, along with tau neutrinos in the $\tau \rightarrow \mu$ decay channel.

Finally, another detector will be installed. Labeled as "Near" detector, it will be placed in pseudo-rapidity range overlapping with LHCb detector, thus reducing the systematic uncertainty of SND@LHC measurements of charm production. The expected high statistics and detector accuracy will thus allow to perform precise measurements of neutrino physics at colliders.

\end{document}